\documentclass{aip-cp}
\usepackage[sort&compress,numbers,sort,elide]{natbib}
\usepackage{rotating}
\usepackage{graphicx}

\begin{document}

\title{Status and Prospects of Discrete Symmetries Tests in Positronium Decays with the J-PET Detector}

\author{M.~Silarski for the J-PET Collaboration}

\affil{Faculty of Physics, Astronomy and Applied Computer Science, Jagiellonian University, S. {\L}ojasiewicza 11, 30-348 Krak{\'o}w, Poland}
\corresp{Corresponding author: michal.silarski@uj.edu.pl}

\maketitle

\begin{abstract}
Positronium is a unique laboratory to study fundamental symmetries in the Standard Model,
reflection in space ($\mathcal{P}$), reversal in time ($\mathcal{T}$), charge conjugation ($\mathcal{C}$)
and their combinations. The experimental limits on the $\mathcal{C}$, $\mathcal{CP}$ and $\mathcal{CPT}$
symmetries violation in the decays of positronium are still several orders of magnitude higher
than the expectations. 
The newly constructed Jagiellonian Positron Emission Tomograph (J-PET)
was optimized for the registration of photons from the electron-positron annihilations. It enables
tests of discrete symmetries in decays of positronium atoms via the determination of the expectation 
values of the discrete-symmetries-odd operators. 
In this article we present the capabilities of
the J-PET detector in improving the current 
precision of discrete symmetries tests
and report on the progress of analysis data from the first 
data-taking runs.
\end{abstract}
\section{INTRODUCTION}
One of the greatest achievements of the twentieth century theoretical physics was a formulation
and proof of Noether's theorem which connects 
symmetries of physical systems
and conservation laws~\cite{noether}. Ever since then symmetries have become an essential toolbox
of almost all physics theories and models, especially in particle physics.
Every quantum field theory describing the interaction and properties of elementary
particles is formulated requiring Lorentz invariance. 
An important role in the Standard
Model formulation has been played by the discrete symmetries of Parity $\mathcal{P}$,
Charge Conjugation $\mathcal{C}$ and Time Reversal $\mathcal{T}$ and their combinations, $\mathcal{CP}$
and $\mathcal{CPT}$.
They prove to be very useful in the calculation of the cross sections and decay rates, especially
for the processes governed by the strong interaction~\cite{michalphd}.
Among all the known forces only the weak interaction 
leads to phenomena violating the invariance under
the three discrete operations and their combination, $\mathcal{CP}$. The $\mathcal{P}$ invariance violation was first 
observed in the $\beta$ decay of $^{60}$Co isotope~\cite{Wu}, while soon after this discovery
it was shown that the $\mathcal{P}$ and $\mathcal{C}$ symmetries are violated in the subsequent decays of charged pions
and muons~\cite{Garwin}. The time reversal symmetry breaking have been observed far only by
the CPLEAR Collaboration in neutral kaons transitions of $\mathcal{CP}$-conjugate states~\cite{K2}
and by the BaBar Collaboration in B mesons decays using transitions between their pure flavour
and $\mathcal{CP}$-definite states~\cite{B9}.
Currently analogous test have been performed for neutral kaons by the KLOE-2
Collaboration~\cite{Gajos,alek,Bernabeu1}.
The combined symmetry $\mathcal{CP}$ operation was thought to be exact
until the regeneration studies of the neutral $K$ mesons by Christenson, Cronin, Fitch and
Turlay~\cite{Christensen}. In the Standard Model the $\mathcal{CP}$ violation mechanism is introduced
by the quark mixing described by the complex Cabibbo-Kobayashi-Maskawa matrix with one nonzero
phase~\cite{cabibo,ckm}, which explicitly requires existence of the three generations of quarks~\cite{michalphd}.
The $\mathcal{CP}$ violation in the kaon sector is well known due to several experiments, mainly the NA48~\cite{na48},
KTEV~\cite{ktev} and KLOE~\cite{passeri}, but there are still several open issues under investigation
mainly related to rare and ultra rare kaon decays. The $\mathcal{CP}$-violating $K_L \to \pi^0\nu\nu$ and
$K^+ \to \pi^+\nu\nu$ decays have been searched recently by the KOTO~\cite{koto} and NA62~\cite{na62}
experiments. The $\mathcal{CP}$ violation is also still not well measured for the $K_S$ meson, especially
in the three-pion and semileptonic decays~\cite{Michal,DariaHEP}.
In the $B$ meson decays the violation of this symmetry appears to be even stronger than for
K mesons~\cite{pdg2018}. As in case of kaons it was found in the mixing of $B^{0}-\bar{B}^0$~\cite{B1,B2}
and directly in the decay amplitudes of $B^{0}$~\cite{B3,B4}, $B^{+}$~\cite{B5,B6,B7} and $B_{S}^{0}$~\cite{B8}.\\ 
The $\mathcal{CP}$ symmetry is very important in view of the observed matter-antimatter asymmetry
which requires much larger violation than predicted by the Standard Model.
Therefore, a lot of effort is made nowadays to find a new sources of $\mathcal{CP}$ and $\mathcal{C}$
symmetries breaking. Apart from the $D$ meson decay studies~\cite{pdg2018} discrete symmetries
violation is searched in baryonic systems~\cite{LambdaB} and in the leptonic sector, in particular
in the positronium decays~\cite{Kaminska,Moskal:2016moj,Yamazaki:2009hp} and in neutrino
oscillations~\cite{Abe:2017vif}.
Invariance under the $\mathcal{CPT}$ operator is also of a great importance since its violation
would be an unambiguous sign of phenomena not included in the Standard Model. There have been performed
many tests, including neutral kaons~\cite{DariaHEP,Adler:1995xy,Ambrosino:2006ek,DeSantis,Ambrosino:2006vr},
B meson system~\cite{Aaij:2016mos} and positronium decays~\cite{cpt-ee},
and so far this symmetry seems to be exact for all the known interactions. 
\section{POSITRONIUM AS A PROBE FOR DISCRETE SYMMETRIES TESTS}
In the leptonic sector positronium is a very promising system to test discrete symmetries and to
look for physics beyond Standard Model. 
It is a simple bound state of an electron and a positron which
formation and decay is governed by QED.
%
%
Positronium is an eigenstate of the $\mathcal{P}$
operator, as a system bound by a central potential, and $\mathcal{C}$ as well
(particle-antiparticle state). 
In the ground state the parity of positronium is equal to
$\lambda_P = -1$, while the charge conjugation eigenvalue depends on spin $S$
of the system: $\lambda_C = (-1)^S$. Thus, para-positronium ($S=0$) is a $\mathcal{CP}$-odd state
and the spin-triplet state, ortho-positronium, is even with respect to this operator.
Since QED is invariant under all the mentioned operations discovery of violation of any discrete
symmetries would be a sign of some new processes not described by the Standard Model.
In case of positronium these tests can be made with a very high precision. For example,
the expected $\mathcal{CP}$ violation effects in the ortho-positronium ($o\textrm{-}Ps$) decays are expected to
be at the level of 10$^{-10}$-10$^{-9}$ due to the photon-photon interactions~\cite{sozzi}.
The tests of invariance under a certain operation may be conducted directly, by searching
of transitions leading to final state with opposite eigenvalue than the initial one.
With positronium one can test in this way for example the $C$ symmetry by searching for $o\textrm{-}Ps$ and $p\textrm{-}Ps$
decays to an even and odd number of photons, respectively. Experiments done so far led to the following
upper limits on branching ratios for both spin states:
$BR\left(o\textrm{-}Ps \to 4\gamma/o\textrm{-}Ps \to 3\gamma\right) < 2.6 \cdot 10^{-6}$ at~90\%~C. L.~\cite{ops}, 
$BR\left(p\textrm{-}Ps \to 3\gamma/p\textrm{-}Ps \to 2\gamma\right) < 2.8 \cdot 10^{-6}$ at~68\%~C. L.~\cite{pps1},
$BR\left(p\textrm{-}Ps \to 5\gamma/p\textrm{-}Ps \to 2\gamma\right) < 2.7 \cdot 10^{-7}$ at~90\%~C. L.~\cite{pps2}.
The alternative way of testing are measurements of non-vanishing
expectation values of certain operators odd under the transformation~\cite{alek1}.
For $o\textrm{-}Ps \to 3\gamma$ they can be constructed from momenta of the final-state photons, their
polarization and spin of the positronium~\cite{Moskal:2016moj}. Some of the operators of interest
are listed in Tab.~\ref{tab1}. 
\begin{table}[h]
\caption{List of operators with non-zero expectation values odd under different
symmetries~\cite{Moskal:2016moj}.}
\label{tab1}
\tabcolsep7pt\begin{tabular}{lll}
\hline
\tch{1}{r}{b}{Operators} & & \tch{1}{l}{b}{Tested Symmetries}\\
\hline
$\vec{S}\cdot \vec{k_1}$ & $\vec{S}\cdot\left( \vec{k_2} \times \vec{\epsilon_1} \right)$ & $\mathcal{P, CP, CPT}$\\
$\vec{S}\cdot\left( \vec{k_1} \times \vec{k_2} \right)$ & $\vec{S}\cdot \vec{\epsilon_1}$ &$\mathcal{T, CPT}$\\
$\vec{S}\cdot \vec{k_1}\left[\vec{S}\cdot\left( \vec{k_1} \times \vec{k_2} \right)\right]$ & $\vec{k_1}\cdot \vec{\epsilon_2}$& $\mathcal{P, T, CP}$\\
\hline
\end{tabular}
\end{table}
The~$\vec{k_1}$, $\vec{k_2}$, $\vec{k_3}$ denote momenta of the three photons in ascending order
according to their moduli~\cite{Moskal:2016moj} and $\epsilon_1$, $\epsilon_2$ and $\epsilon_3$
are their polarization vectors, respectively. The quantization axis of the spin of $o\textrm{-}Ps$
requires a static magnetic field, which mixes the two positronium spin states for the spin
projection $j_z = 0$. 
This shortens effectively the ortho-positronium lifetime and allows one to
determine the spin polarization. For elemental positron sources one can
determine the spin direction taking advantage that due to the parity violation the positrons
are linearly polarized along their velocity direction. This polarization is preserved to high
extend during the positronium formation~\cite{Moskal:2016moj,Coleman}.
Up to now the experiments were 
performed only for the $\mathcal{CP}$
and $\mathcal{CPT}$ symmetries yielding the following mean values of corresponding
operators: $\left\langle \vec{S}\cdot \vec{k_1}\left[\vec{S}\cdot\left( \vec{k_1} \times \vec{k_2} \right)\right]\right\rangle
= 0.0013 \pm 0.0012$~\cite{Yamazaki:2009hp}
and $\left\langle \vec{S}\cdot\left( \vec{k_1} \times \vec{k_2} \right)\
\right\rangle = 0.0071 \pm 0.0062$~\cite{cpt-ee}. 
\section{PROSPECTS OF DISCRETE SYMMETRIES TESTS WITH THE J-PET DETECTOR}
\begin{figure}[h]
 \includegraphics[width=0.28\textwidth]{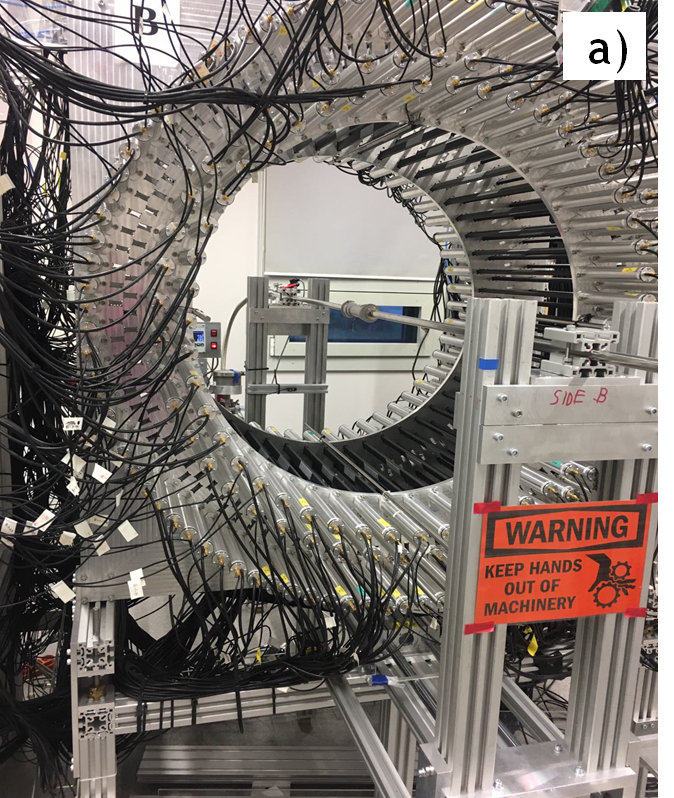}
 \includegraphics[width=0.32\textwidth]{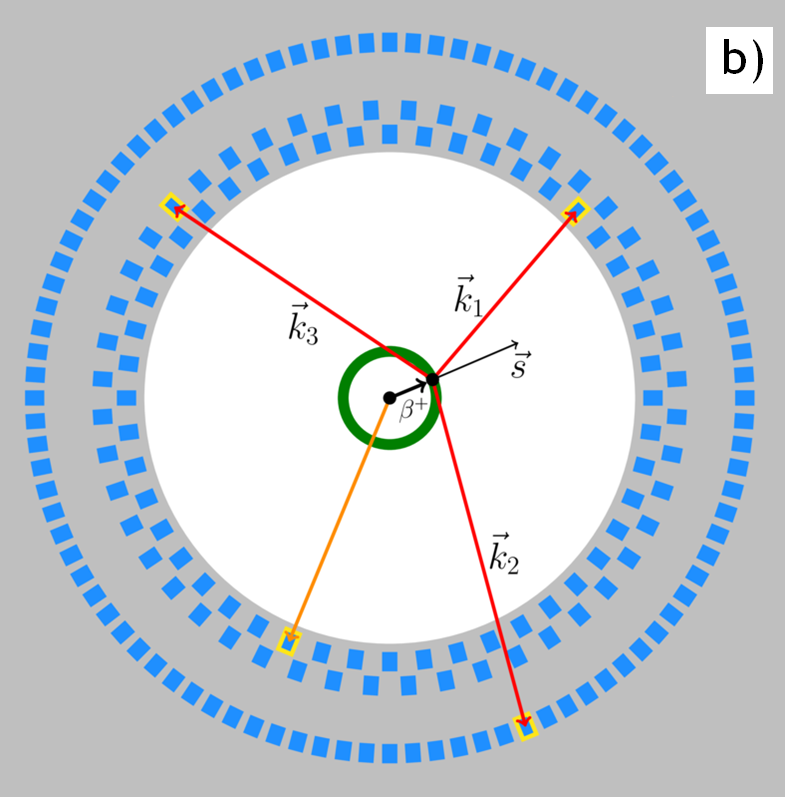}
  \caption{ a) Photograph of the J-PET detector with a vacuum chamber installed.
	b) Schematic representation of the principle of ortho-positronium decays studies with the J-PET
	detector\cite{Moskal:2016moj}. An open $^{22}$Na source is placed inside a vacuum chamber (red circle)
	in the geometric
	center of the J-PET detector. Positron originating from the $\beta^+$ decay may form ortho-positronium
	in a porous material placed on the inner surface of the chamber (black dot). The three gamma quanta
	from $o\textrm{-}Ps \to 3\gamma$ decay, with momenta $\vec{k_1}$, $\vec{k_2}$, $\vec{k_3}$, are registered
	by the J-PET detector. The measurement of $o\textrm{-}Ps$ lifetime is provided by the registration of the
	gamma quantum from the deexcitation of the $^{22}$Ne isotope originating from the sodium $\beta^+$
	decay. Figure adapted from~\cite{Mohammed:2017dkc}.}
	\label{fig1}
\end{figure}

One of the ongoing experiments having potential to significantly improve the sensitivity of
the discrete symmetries is the J-PET experiment~\cite{Kaminska,Moskal:2016moj,Moskal:2014sra,Moskal:2014rja,trilateration}.
Designed as novel cost-effective and full-body PET scanner~\cite{Niedzwiecki:2017nka,Moskal:2016ztv,kowalski},
the J-PET detector constitutes a multipurpose device for fundamental particle physics studies~\cite{Moskal:2016moj,nature,Nowakowski}.
In its current configuration it consists of 192 detection modules made out of long EJ-230 scintillator bars with dimensions
19x7x500~mm$^3$ arranged in three layers. Light produced by gamma quanta interaction in each
scintillator is collected by Hamamatsu R9800 photomultipliers connected optically to both
ends of the strip. The J-PET detection setup is presented in Fig.~\ref{fig1}a). Photomultipliers
signals are probed at four thresholds on both, leading and trailing edges by novel frond-end
electronics with accuracy of about 30~ps~\cite{Palka}. It provides also signal's
charge determination by the Time-Over-Threshold (TOT) measurement, which leads to the determination
of the gamma quanta energy loss with accuracy of about $\sigma(E)/E = 0.044/E(\mathrm{MeV})$~\cite{Moskal:2014sra}.
The J-PET FPGA-based data acquisition system is working in a continuous readout mode and allow true
real-time tomographic data processing~\cite{Korcyl1,Korcyl2}. The J-PET time resolution
of annihilation gamma quanta registration amounts to about $\sigma_t \sim 125$~ps~\cite{Moskal:2014sra}
and it will be further improved by using dedicated reconstruction methods taking advantage of signals
probing at the four thresholds~\cite{Raczynski:2014poa,Moskal:2014rja,Sharma:2015rza}.
Using the J-PET detector one can test the discrete symmetries both, by searching for forbidden
decays and by measurement of symmetry-odd operator expectation values. As it was shown in Fig.~\ref{fig1}b{)
the measurements have been done with a $^{22}$Na positron source, enclosed in a thin kapton
foil to minimize annihilations in the source itself, placed in an vacuum chamber covered with
a material maximizing the formation of $o\textrm{-}Ps$ with a long lifetime~\cite{Jasinska:2016qsf}.
As it was mentioned in the previous
section the positron is linearly polarized along its velocity and there is a high probability
that the spin direction does not change during the $o\textrm{-}Ps$ formation. Thus, to determine the ortho-positronium
spin direction one can measure the positron velocity direction. This can be done by reconstructing the
position of the positronium formation by a dedicated trilateration method~\cite{trilateration}.
%
%
The $o\textrm{-}Ps$ lifetime measurement is done using a gamma quantum from excited $^{22}$Ne
associated to the positron emission. It can be distinguished from the annihilation and scattered photons
via the TOT measurement which is much higher for deexcitation quanta. The three photons originating
from the $o\textrm{-}Ps$ decay are detected with an efficiency of about $10^{-5}$-$10^{-4}$ depending on the energy
loss threshold used~\cite{Kaminska}.
The plastic scintillators used in the J-PET detector enable also determination for the gamma quanta
polarization. Since the most probable effect of annihilation gamma quanta interaction is the Compton
scattering occurring most likely in the plane perpendicular to the electric vector of the photon,
measurement of four-momenta of annihilation and corresponding scattered quanta allows to determine,
to some extend, linear polarization of the primary quanta, e.g. by the following cross product:
$\vec{\epsilon_\gamma} = \vec{k} \times \vec{k'}$, where $\vec{k}$ and $\vec{k'}$ are the momenta of
the annihilation and scattered photon~\cite{Moskal:2016moj}.
The J-PET detector have been successfully commissioned and have been taking data with several different
vacuum chambers and targets for $o\textrm{-}Ps$ formation~\cite{Czerwinski,Skurzok}. The preliminary analysis
of gathered data revealed that the main background for the ortho-positronium decay studies
is composed by the pick-off annihilations or ortho-para spin conversion due to the spin-orbit
interaction or due to electron exchange~\cite{Moskal:2016moj}. These processes lead to the final
state with two 511~keV photons which may scatter and mimic the $o\textrm{-}Ps \to 3\gamma$ decay.
This background can be reduced by considering the distribution of the sum and difference of the two
smallest relative angles between reconstructed directions of photons (see Fig.~\ref{fig2}a) )~\cite{Kaminska}.
\begin{figure}[ht]
 \includegraphics[width=0.45\textwidth]{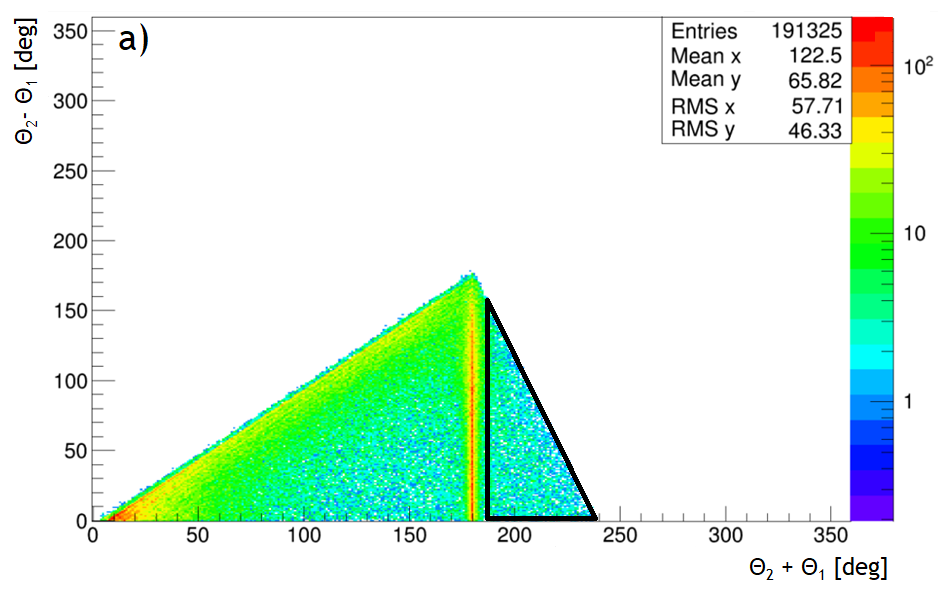}
 \includegraphics[width=0.47\textwidth]{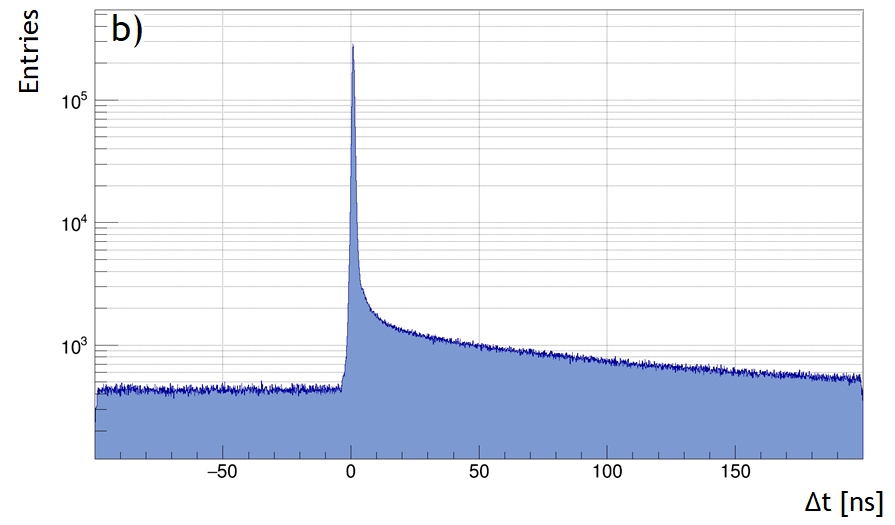}
  \caption{ a) Distribution of the difference of two smallest relative angles between
	reconstructed momenta of photons in a function of their sum obtained with a small sub-sample
	of the J-PET data. The $o\textrm{-}Ps \to 3\gamma$ signal region is showed as a black triangle; b) Lifetime
	spectrum of positronium registered with the J-PET detectors.}
	\label{fig2}
\end{figure}
In case of the para-positronium forbidden decays, e.g. 
$p\textrm{-}Ps \to 3\gamma$, the background originating
from the corresponding ortho-positronium one can take advantage of the completely different
lifetimes of the two spin states. They can be separated using the time difference $\Delta t$ between the
registration of annihilation photons and deexcitation gamma quantum. An exemplary distribution of
such difference measured with the J-PET detector is shown in Fig.~\ref{fig2}b). Such distributions can be
used also in the Positron Annihilation Lifetime measurements which may be used in medical
diagnostics~\cite{Jasinska:2017ngd,Kubicz:2015nga}.
\section{SUMMARY}
Discrete symmetries have been playing an exceptional role in formulation and tests of the Standard Model.
Violation of the $\mathcal{P}$, $\mathcal{C}$ and $\mathcal{CP}$ symmetries by the weak interactions
is well established for neutral kaons
and $B$ mesons, while there is still no experimental evidence of analogous breaking in the baryonic
or leptonic sectors. 
As the lightest purely charged leptonic state, 
positronium constitutes one of the best systems
to search for new effects not included in the Standard Model. In this case the symmetries can be tested
by searching for the forbidden decays, e.g. $p\textrm{-}Ps \to 3\gamma$, or by measurement of expectation values
of symmetry-odd operators. One of the experiments which has a great potential in providing new experimental
data in this sector is the J-PET detector since it has a unique capability of positronium spin determination
and gamma quanta polarization~\cite{ksiazka}. With the present detector and near future
upgrades~\cite{Moskal:2016ztv,Moskal:2018wfc,Raczynski:2017rfm,totpet} 
J-PET will be able
to test discrete symmetries in the charged lepton system
with significantly higher sensitivity than presently published results.    
\section{ACKNOWLEDGMENTS}
The author would like to express his gratitude to prof. Steven Bass for proof-reading of the article and many useful comments.
We acknowledge technical and administrative support of A. Heczko, M. Kajetanowicz
and W. Migda{\l}. This work was supported by The Polish National Center for Research
and Development through grant INNOTECH-K1/IN1/64/159174/NCBR/12, the Foundation
for Polish Science through the MPD and TEAM/2017-4/39 programmes, the National Science
Centre of Poland through grants no. 2016/21/B/ST2/01222, 2017/25/N/NZ1/00861, the
Ministry for Science and Higher Education through grants no. 6673/IA/SP/2016, 7150/E338/SPUB/2017/1
and 7150/E-338/M/2017, and the Austrian Science Fund FWF-P26783.
\nocite{*}
\bibliographystyle{aipnum-cp}%

\end{document}